\documentclass[preprint,notoc]{JHEP3}
\usepackage{amssymb,amsmath,bbm,bm,dsfont}
\usepackage{epsfig}
\usepackage{graphicx}
\usepackage{citesort}
\DeclareOldFontCommand{\rm}{\normalfont\rmfamily}{\mathrm}
\DeclareOldFontCommand{\bf}{\normalfont\bfseries}{\mathbf}

\def\bfn{{\bf f}_{\nu}}

\def\bcm2{\bm{\mathcal{M}}^2}
\def\ve{\varepsilon}
\def\a{\alpha}

\def\k{\kappa}

\def\p{\pi}

\def\G{\Gamma}

\newcommand{\beq}{\begin{equation}}
\newcommand{\eeq}{\end{equation}}
\def\bea{\begin{eqnarray}}
\def\eea{\end{eqnarray}}

\title{Examining leptogenesis with lepton flavor violation and the dark matter abundance}
\author{Steve Blanchet$^a$, Danny Marfatia$^b$, Azar Mustafayev$^c$\\
        $^a$Department of Physics, University of Maryland, College Park, MD 20742, USA\\
        $^b$Department of Physics \& Astronomy, University of Kansas, Lawrence, KS 66045, USA\\
        $^c$William I.~Fine Theoretical Physics Institute, University of Minnesota,
        Minneapolis, MN 55455, USA\\
        E-mail: \email{sblanche@umd.edu}, \email{marfatia@ku.edu},
        \email{mustafayev@physics.umn.edu}}

\preprint{UMN--TH--2907/10\\
        FTPI--MINN--10/15\\
        UMD-PP-10-009}

\abstract{ Within a supersymmetric (SUSY) type-I seesaw framework
with flavor-blind universal boundary conditions, we study the
consequences of requiring that the observed baryon asymmetry of the Universe be
explained by either thermal or non-thermal leptogenesis. In the
former case, we find that the parameter space is very constrained. 
In the bulk and stop-coannihilation regions of mSUGRA parameter space (that are 
consistent with the measured dark matter abundance),
lepton flavor-violating (LFV) processes are accessible at MEG 
and future experiments. However, the very high
reheat temperature of the Universe needed after inflation (of about $10^{12}$~GeV) 
leads to a severe gravitino problem, which disfavors either thermal leptogenesis 
or neutralino dark matter.
Non-thermal leptogenesis in the preheating phase from SUSY flat
directions relaxes the gravitino problem by lowering the required reheat
temperature. The baryon asymmetry can then be
explained while preserving neutralino dark matter, and for the
bulk or stop-coannihilation regions LFV processes should be observed in current or
future experiments.
 }

\begin{document}

\section{Introduction}

Supersymmetry is perhaps the leading possibility for physics beyond
the Standard Model. One of its nice features is that it contains
natural candidates for the observed dark matter in the Universe.
Within the supersymmetric model with minimal particle content
(MSSM), it is customary to assume flavor-blind boundary conditions
at the Grand Unified Theory (GUT) scale, in which case the model is
referred to as mSUGRA or constrained MSSM (CMSSM). If the dark matter particle is the
lightest neutralino, the parameter space of
mSUGRA is very tightly constrained by the precisely determined dark
matter abundance in the Universe~\cite{Komatsu:2008hk}.

With exact R-parity conservation, neutrinos are
massless in mSUGRA. However, there is now overwhelming evidence that
neutrinos have mass and mix; for a review see Ref.~\cite{rev}. 
The simplest explanation for small
neutrino masses is perhaps the type-I seesaw
mechanism~\cite{seesaw1}. It is therefore natural to extend mSUGRA
to allow for a seesaw mechanism and thus for small neutrino masses.
An mSUGRA-seesaw with $SO(10)$-inspired boundary conditions was
recently studied in~\cite{Barger:2008nd}, where it was found that
neutrinos, with their Yukawa couplings contributing to the running
of various parameters, such as the slepton mass matrices and the
trilinear couplings, substantially modify the parameter space allowed by dark
matter. Lepton flavor violation (LFV) was then
studied within the same framework, and the LFV rates were shown to
potentially differ from existing estimates by up to two orders
of magnitude~\cite{Barger:2009gc}.

In this paper we add yet another constraint to the
mSUGRA-seesaw+dark-matter scenario, namely that the
baryon asymmetry of the Universe be explained by either thermal or
non-thermal leptogenesis.

It is well-known that in $SO(10)$-inspired scenarios where the
type-I seesaw mechanism provides the dominant contribution to
neutrino masses, thermal leptogenesis~\cite{Fukugita:1986hr}
typically fails to explain the observed baryon asymmetry of the
Universe. The reason is that the lightest right-handed (RH) neutrino
is generally too light to generate enough
asymmetry~\cite{Akhmedov:2003dg}. Including flavor
effects~\cite{Barbieri:1999ma,Abada:2006fw}, the
situation improves since the next-to-lightest RH
neutrinos (not accounted for in Ref.~\cite{Akhmedov:2003dg}), can generate a large
asymmetry~\cite{DiBari:2005st}. Nevertheless, the scenario remains
tightly constrained~\cite{Abada:2008gs,DiBari:2008mp}, which
perfectly suits our purpose: If thermal leptogenesis is successful
only in a very restricted part of the parameter space, then
definite predictions for LFV rates at a given point in the
mSUGRA parameter space are possible. 

Thermal leptogenesis with hierarchical RH neutrinos requires the
reheat temperature after inflation to be above
$10^9$~GeV~\cite{Davidson:2002qv,Giudice:2003jh,Buchmuller:2004nz}.
In mSUGRA this poses a problem because of the overproduction of
gravitinos~\cite{gravitino}. This tension is partially alleviated if
the gravitino is heavier than 30 TeV, as the reheat temperature
is then allowed to be as high as
$10^{10}$~GeV~\cite{Kawasaki:2008qe}. As we show,
such a reheat temperature is not high enough to allow
for thermal leptogenesis from the next-to-lightest RH neutrino
decays. The consequence is that thermal leptogenesis in our mSUGRA
$SO(10)$-inspired framework is inconsistent with neutralino dark
matter (or even gravitino dark matter).

The alternative possibility of non-thermal
leptogenesis, either at reheating from inflaton
decay~\cite{Asaka:1999jb,Asaka:1999yd}, or at
preheating~\cite{Giudice:1999fb} allows for lower reheat temperatures than thermal
leptogenesis. We employ the mechanism of instant
preheating~\cite{Felder:1998vq} from SUSY flat directions, as
presented in~\cite{Giudice:2008gu}. We show that this mechanism is able to
successfully explain the baryon asymmetry of the Universe, while
maintaining the viability of the neutralino or the gravitino as dark matter candidates.
Our predictions for LFV rates turn out to be close to the current bounds
for mSUGRA points in the bulk region.

It is worth mentioning that a mixed type-I + type-II~\cite{seesaw2}
seesaw mechanism can be naturally obtained within $SO(10)$, and
leptogenesis becomes much
easier~\cite{Hosteins:2006ja,Abada:2008gs}. However, for the sake of
minimalism and the predictiveness, we limit ourselves to a
dominant type-I case only.

In Section 2 we describe the framework in which we
 work. In Section 3 we review thermal leptogenesis,
introducing all the necessary tools for our computation. We also
show the numerical results for the predicted LFV rates, and comment
on the gravitino problem. In Section 4 we perform the same analysis
with non-thermal leptogenesis at preheating. In Section 5 we
summarize our findings and conclude.

\section{SUSY-seesaw and SO(10) GUTs}
We consider the following superpotential for the MSSM augmented by
singlet right-handed neutrinos $\hat{N}^c_i$: \beq
\hat{f}=\hat{f}_{\rm MSSM}+ (\bfn)_{\alpha
j}\epsilon_{ab}\hat{L}^a_{\alpha}\hat{H}_u^b\hat{N}^c_i +
\frac{1}{2}({\bf M}_N)_{ij}\hat{N}^c_i \hat{N}^c_j \, ,
\label{eq:RHNspot} \eeq 
where $\alpha$ is the lepton flavor index, $i,j$ are generation indices, $a,b$ are 
$SU(2)_L$ doublet indices, $\epsilon_{ab}$ is the totally antisymmetric tensor with 
$\epsilon_{12}=1$, and the superscript $c$ denotes charge conjugation.
Here, $\hat{f}_{\rm MSSM}$ is the MSSM
superpotential, $\hat{L}$ and $\hat{H}_u$ are, respectively, the
lepton doublet and up-Higgs superfields, and ${\bf M}_N$ is
the Majorana mass matrix for the (heavy) right-handed neutrinos. At
energy scales above ${\bf M}_N$, the light neutrino mass matrix is
given by the type-I seesaw formula~\cite{seesaw1}, \beq
\mathcal{M}_{\nu} =-\bfn {\bf M}_N^{-1} \bfn^T v^2_u \, ,
\label{eq:seesaw} \eeq where $v_u$ is the vacuum expectation value
(VEV) of the neutral component $h^0_u$ of the up-type Higgs doublet
$H_u$. We denote the eigenvalues of the light neutrino mass matrix by
 $m_{\nu i}$, $i=1,2,3$. In the limit in which all RH neutrinos
are decoupled, the light neutrino mass matrix is
$\mathcal{M}_{\nu}=-\kappa v^2_u$; $\kappa$ is the
coupling matrix of the dimension-5 effective operator generated by
RH neutrinos, which is determined by matching conditions at the RH
neutrino decoupling thresholds. The matrix $\mathcal{M}_{\nu}$ is
diagonalized (in the basis where charged leptons are diagonal) by
the MNS matrix that can be parameterized by the mixing angles
$\theta_{12},\ \theta_{23}$, and $\theta_{13}$, the Dirac phase
$\delta$, and two Majorana phases $\phi_1$ and
$\phi_2$ (see Ref.~\cite{Barger:2009gc} for our convention).

Inspired by $SO(10)$ GUTs, we introduce the vector $R_{\nu u}=(R_1,R_2,R_3)$, 
with strictly positive entries, which relates the diagonal up-type quark Yukawa couplings ${\bf
f}_u^{\rm diag}$ to the diagonal neutrino Yukawa couplings ${\bf
f}_{\nu}^{\rm diag}$ at the GUT scale:
\begin{equation}\label{quarkneutrino}
\left({\bf f}_{\nu}^{\rm diag}\right)_{ij} = R_{i} \left({\bf
f}_u^{\rm diag}\right)_{ij}\,.
\end{equation}
Since we will always assume $R_i<\mathcal{O}(5)$, a highly
hierarchical pattern ${\bf f}_{\nu 3}^{\rm diag}\gg {\bf f}_{\nu
2}^{\rm diag}\gg {\bf f}_{\nu 1}^{\rm diag}$ is obtained. Note that
the $R_i$'s are in general all different, and in the minimal
$SO(10)$ scenario, the range is typically $1\leq R_i\leq 3$. However, higher
values $R_i\gtrsim 5$ can be easily achieved with non-renormalizable
operators. The condition in Eq.~(\ref{quarkneutrino}) corresponds to
an extension of the ``small mixing'' scenario presented
in~\cite{Barger:2009gc}. We will not consider the ``large mixing''
case here, since large regions of the parameter space are already
excluded by existing bounds on $\tau \to \mu\gamma$~\cite{Barger:2009gc}.

Equation~(\ref{quarkneutrino}) implies that the RH
neutrinos have a very strong hierarchy. To see this, assume tribimaximal mixing for
the light neutrinos, and neglect the small CKM-type mixing in ${\bf
f}_{\nu}$. For a normal hierarchy of light neutrinos, $m_{\nu 1}\ll
m_{\nu 2} \ll m_{\nu 3}$, one obtains~\cite{Barger:2009gc}:
\begin{equation}
 M_{N_1}\simeq\frac{3m_u^2}{m_{\nu_2}}R_{1}^2\,,\
 M_{N_2}\simeq\frac{2m_c^2}{m_{\nu_3}}R_{2}^2\,,\
 M_{N_3}\simeq\frac{m_t^2}{6m_{\nu_1}}R_{3}^2\,,
\label{eq:ckmRHN}
\end{equation}
whereas for the inverted mass
hierarchy ($m_{\nu_1}\simeq m_{\nu_2} \gg m_{\nu_3}$),
\begin{equation}
 M_{N_1}\simeq\frac{3m_u^2}{m_{\nu_2}}R_{1}^2\,,\
 M_{N_2}\simeq\frac{2m_c^2}{3m_{\nu_1}}R_{2}^2\,,\
 M_{N_3}\simeq\frac{m_t^2}{2m_{\nu_3}}R_{3}^2\,.
\label{eq:ckmRHN2}
\end{equation}
From the above scaling behavior, we see that a quasi-degenerate
spectrum ($m_{\nu_1}\simeq m_{\nu_2}\simeq m_{\nu_3}$) would require
the lightest Majorana mass to be in the $10^2$--$10^3$~GeV range
with significant L-R mixing in the sneutrino sector, which we
disregard because it would substantially complicate the sneutrino mass
spectrum and phenomenology. Moreover, since the next-to-lightest RH
neutrino is also lighter than in the case of the normal hierarchy,
successful thermal leptogenesis is rendered more difficult.
The inverse hierarchical case would require the heaviest Majorana
mass to be of order $10^{17}$~GeV, which is well above the GUT
scale. This type of spectrum also suffers from instabilities under
very small changes to ${\bf M}_N$ and RGE
evolution~\cite{Albright:2004kb}. For all these reasons we choose to
focus on the normal hierarchy of 
light neutrinos, $m_{\nu 1}\ll m_{\nu 2} \ll m_{\nu 3}$.

\section{Thermal leptogenesis}

Thermal leptogenesis is one of the most popular mechanisms to
explain the observed baryon asymmetry of the
Universe~\cite{Fukugita:1986hr}. The crucial parameters for
leptogenesis are the $C\!P$ asymmetries $\tilde{\ve}_{i\a}$ and the
washout parameters $K_{i\alpha}$.  The $C\!P$ asymmetry from the
decay of the heavy (s)neutrino $N_i$ ($\tilde{N}_i$) into a (s)lepton of flavor
$\alpha$ is in full generality given
by~\cite{Covi:1996wh}
\begin{equation}
\tilde{\ve}_{i\a}=\frac{1}{8 \p (\bfn^{\dag}\bfn)_{ii}} \sum_{j\neq
i} \left\{ {\rm Im}\left[(\bfn^\star)_{\a i} (\bfn)_{\a j}(\bfn^{\dag}\bfn)_{i j}\right] g(x_j/x_i)+
\frac{2}{(x_j/x_i-1)}{\rm Im} \left[(\bfn^\star)_{\a i}(\bfn)_{\a j}(\bfn^{\dag}\bfn)_{j i}\right]\right\} ,\label{CPasym}
\end{equation}
where $x_i\equiv M_{N_i}^2/M_{N_1}^2$ and \beq\label{xi} g(x)=
\sqrt{x} \left[{2\over x-1}+\ln\left({1+x\over x}\right)\right] \, .
\eeq 
In the SUSY limit, the decay width is
\begin{eqnarray}
\G(N_i\to \ell_{\a}H_u)+\G(N_i\to\bar{\ell}_{\alpha}H_u^{\dagger})&=&\G(N_i\to
\tilde{\ell}_{\a}\tilde{H}_u)+\G(N_i\to \tilde{\ell}^*_{\alpha}(\tilde{H}_u)^c)\nonumber
\\
&=&\G(\tilde{N}^*_i\to \ell_{\alpha}\tilde{H}_u)=\G(\tilde{N}_i\to
\tilde{\ell}_{\alpha}H_u)={|(\bfn)_{\a i}|^2\over 8\pi}
M_{N_i}.\nonumber
\end{eqnarray}
We can then define \beq K_{i\alpha}\equiv
{\G(N_i\to \ell_{\a}H_u)+\G(N_i\to\bar{\ell}_{\alpha}H_u^\dagger)\over
H(T=M_{N_i})}={v_u^2\over m_{\star} M_{N_i}}|(\bfn)_{\a i}|^2\,,
\eeq where $m_{\star}\simeq (1.56\times 10^{-3}~{\rm
eV})\sin^2\beta$.

The baryon asymmetry is obtained by solving a set
of coupled Boltzmann equations as given for instance in Ref.~\cite{Abada:2008gs}. 
However, we use convenient semi-analytical
expressions for the final baryon asymmetry.
The quantity that describes how efficiently the asymmetry is
produced is the \emph{efficiency factor} $\k$, which is a function
of $K_{i\alpha}$. For an initial thermal abundance of RH
(s)neutrinos, $\kappa$ is given
by~\cite{Buchmuller:2004nz},\footnote{For 
the SM case, the replacement
$2\,K_{i\alpha}\to K_{i\alpha}$ must be made because of the fewer
decay modes for each heavy particle.}
\begin{equation}\label{kappa}
\k(K_{i\a}) \equiv {1\over K_{i\a}\,z_B(2K_{i\a})}\,
\left[1-\exp\left(-{2 K_{i\a}\,z_B(2K_{i\a})\over 2}\right)\right]
\, ,
\end{equation}
where
\begin{equation}
z_{B}(K) \simeq 2+4\,K^{0.13}\,\exp\left(-{2.5\over K}\right) \, .
\end{equation}
With a vanishing initial abundance of RH (s)neutrinos, a different
result ensues. A fit valid both in the weak washout
($K_{i\alpha}<3$) and in the strong washout regime
($K_{i\alpha}>3$), was obtained in~\cite{Abada:2006ea}: \beq
\label{vanishing}\k(K_{i\alpha})\simeq \left[ \left({2.6\over
K_{i\alpha}}\right)+\left({K_{i\alpha}\over
0.06}\right)^{1.16}\right]^{-1} \,. \eeq

For the hierarchical mass spectrum of the RH (s)neutrinos
$M_{N_1}\ll M_{N_2}\ll M_{N_3}$, the asymmetry production
from each RH (s)neutrino can be considered separately, and
eventually summed to obtain the final asymmetry.
There are three mass ranges that need to be considered: the
three-flavor regime for $M_{N_i}<(1+\tan^2\beta)\times 10^9~{\rm
GeV}$, the two-flavor ($e+\mu$ and $\tau$) regime for
$(1+\tan^2\beta)\times 10^9~{\rm GeV}<M_{N_i}< (1+\tan^2\beta)\times
10^{12}~{\rm GeV}$ and the unflavored regime for $M_{N_i}>
(1+\tan^2\beta)\times 10^{12}~{\rm
GeV}$~\cite{Barbieri:1999ma,Abada:2006fw}.

It is well-known that with $SO(10)$-inspired mass relations such as
in Eq.~(\ref{quarkneutrino}), the lightest RH (s)neutrino
$N_1~(\tilde{N}_1)$ typically fails to produce enough asymmetry because
$M_{N_1}$ is usually predicted to be much smaller
than the Davidson-Ibarra bound~\cite{Davidson:2002qv} of about
$10^9$~GeV for successful leptogenesis. Note however that tiny
regions in the parameter space exist where RH neutrino masses are
quasi-degenerate, $M_{N_1}\simeq M_{N_2}\simeq M_{N_3}$, in which
case the $C\!P$ asymmetry can be dramatically
enhanced~\cite{Flanz:1996fb} and
leptogenesis is possible~\cite{Akhmedov:2003dg}. We do not
entertain this possibility any further.

We consider the crucial contribution to leptogenesis to arise from the
next-to-lightest RH (s)neutrino,
$N_2~(\tilde{N}_2)$~\cite{DiBari:2005st}. The asymmetry
is typically produced in the two-flavor regime, but it is necessary to
include the potential washout from $N_1~(\tilde{N}_1)$, which occurs
in the three-flavor regime. We then
have~\cite{Blanchet:2008pw,DiBari:2008mp} \bea \eta_{B,2}&\simeq&
0.96\times
10^{-2}\left[\,\tilde{\ve}_{2e}\,\kappa(K_{2e}+K_{2\mu})\exp\left(-{3\p\over
4}K_{1e}\right)+
\tilde{\ve}_{2\mu}\,\kappa(K_{2e}+K_{2\mu})\exp\left(-{3\p\over 4}K_{1\mu}\right)\right.\nonumber \\
&&\left.\hspace{4cm}+~\tilde{\ve}_{2\tau}\,\kappa(K_{2\tau})
\exp\left(-{3\p\over 4}K_{1\tau}\right)\right].\label{thermal} \eea
If the asymmetry from $N_2~(\tilde{N}_2)$ is produced in the
three-flavor regime, \bea
\eta_{B,2}&\simeq& 0.96\times
10^{-2}\left[\,\tilde{\ve}_{2e}\,\kappa(K_{2e})\exp\left(-{3\p\over
4}K_{1e}\right)+
\tilde{\ve}_{2\mu}\,\kappa(K_{2\mu})\exp\left(-{3\p\over 4}K_{1\mu}\right)\right.\nonumber \\
&&\left.\hspace{4cm}+~\tilde{\ve}_{2\tau}\,\kappa(K_{2\tau})
\exp\left(-{3\p\over 4}K_{1\tau}\right)\right]. \eea Since the
asymmetry production from the heaviest RH (s)neutrino,
$N_3~(\tilde{N}_3)$, occurs typically in the unflavored regime, the
$C\!P$ asymmetry is suppressed by $(M_{N_{1,2}}/M_{N_3})^2$, and we
neglect it.

The total asymmetry is given by the sum of the three RH neutrino
contributions, of which just one is relevant, and thus \beq
\eta_B\simeq \eta_{B,2}\,, \eeq to be compared with the measured
value~\cite{Komatsu:2008hk} \beq \label{observed}\eta_B^{\rm CMB}=(6.2\pm
0.15)\times 10^{-10}. \eeq

As we shall see in the next subsection, the final baryon asymmetry
produced through leptogenesis will be typically dependent on the
initial abundance of RH (s)neutrinos; Eq.~(\ref{kappa})
is valid for an initial thermal abundance of $N_2~(\tilde{N}_2)$ whereas
Eq.~(\ref{vanishing}) is valid for a vanishing one. Note that within our
$SO(10)$-inspired scenario, a thermal $N_2~(\tilde{N}_2)$-abundance
is very easily obtained if the $Z'$ of $U(1)_{B-L}$, which is
naturally present if $SO(10)$ breaks to the left-right model, is
heavier than $N_2~(\tilde{N}_2)$ by two orders of magnitude (and
$M_{Z'}\lesssim 10\,T_{\rm R}$, with $T_{\rm R}$ the reheat temperature)~\cite{Racker:2008hp}. We then have
that the interaction $\overline{N}_2 \gamma^{\mu}N_2Z'_{\mu}$
efficiently brings the next-to-lightest RH (s)neutrino into
equilibrium without interfering with the production mechanism from
$N_2~(\tilde{N}_2)$ decays.

\subsection{Results} \label{sec:results}

As stated in Section 2, we present results only for the normal
hierarchy of light neutrinos, $m_{\nu 1}\ll m_{\nu 2}\ll m_{\nu 3}$.
More precisely, we require that $m_{\nu 1}<m_{\rm
sol}=0.009$~eV.

We use ISAJET-M to produce the neutrino and SUSY
spectra~\cite{Barger:2009gc}. The program implements RGE
evolution in the MSSM with a type-I seesaw in full matrix form at
the 2-loop level. All sparticle masses are computed with complete
1-loop corrections and gauge and Yukawa coupling evolution include
multiple sparticle threshold effects.  The decoupling of RH
neutrinos is performed at multiple scales equal to their own running
masses, and effects of $\bfn$ on the MSSM parameters are included.
It is known that large $\bfn$ entries can significantly affect the
RGE evolution with concomitant effects on the MSSM spectrum and the
neutralino DM rates~\cite{Barger:2008nd,Barger:2009gc,nuSUGRA}.

In the neutrino sector, we use a top-down approach with $\bfn$
and ${\bf M}_N$ input at $M_{\rm GUT}$ and the neutrino
mass matrix $\mathcal{M}_{\nu}$ obtained by RGE evolution. We fix
$\bfn (M_{\rm GUT})$ using the $SO(10)$-inspired relation
(\ref{quarkneutrino}) and adjust ${\bf M}_N$ to produce a viable
spectrum of light neutrinos. Since leptogenesis is sensitive to
details of the MNS matrix, we scan over 5 parameters:
$\phi_1$, $\phi_2$, $\delta$, 
$\theta_{13}$ and $m_{\nu 1}$.

\begin{figure}
\center
\includegraphics[width=0.85\textwidth]{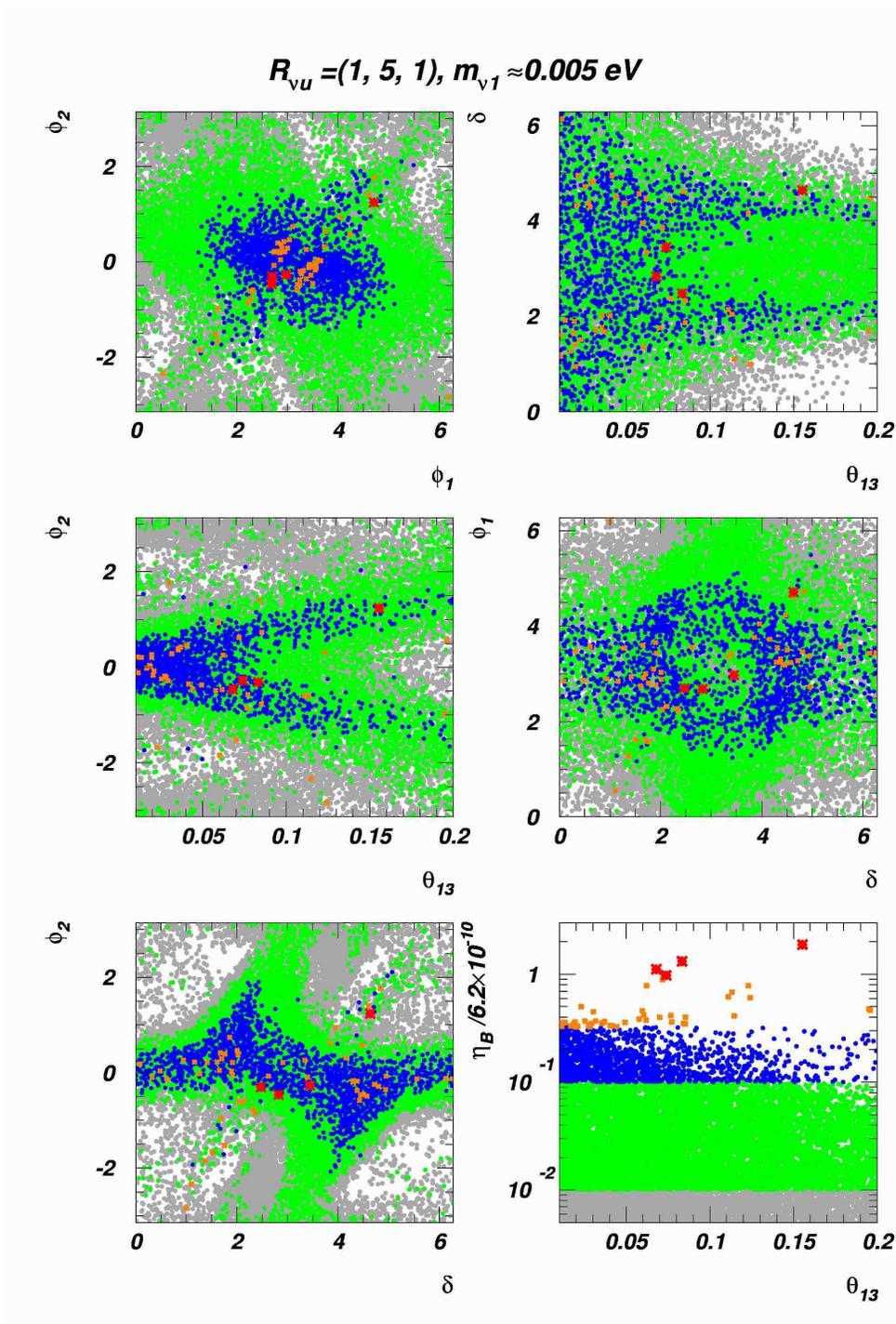}
\caption{Full parameter space scan for thermal leptogenesis (thermal
initial $N_2$-abundance). The color code is evident from the bottom-right
panel.}
\label{fig:1}
\end{figure}

In Fig.~\ref{fig:1}, we show the results for the baryon asymmetry
generated by thermal leptogenesis for $m_{\nu 1}=0.005$~eV and $R_{\nu}=(1,5,1)$ with
 $\phi_1$, $\phi_2$, $\delta$ and
$\theta_{13}$ varied. Note that the figure displays \emph{weak scale} values for all the
parameters.
From the bottom-right panel, we see that there are only a few
points above the $2\sigma$ lower bound on the observed baryon
abundance [cf. Eq.~(\ref{observed})], i.e., $\eta_B>5.9\times
10^{-10}$. This clearly shows that the parameter space that yields
successful leptogenesis is quite restricted.

As noted in Ref.~\cite{DiBari:2008mp}, the baryon asymmetry is essentially
independent of $R_1$, and only mildly dependent on $R_3$. (At the end of the section we
discuss the role of $R_3$ on the predictions for the
LFV rates.) On the other hand,
$R_2$ is crucial for leptogenesis, in that it fixes the
next-to-lightest RH neutrino mass scale [cf. Eq.~(\ref{eq:ckmRHN})],
which itself sets the size of the $C\!P$ asymmetry [cf.
Eq.~(\ref{CPasym})]. We confirm the finding of Ref.~\cite{DiBari:2008mp}
that leptogenesis is only possible for $R_2>3$,
and set $R_2=5$ in our calculations.

\begin{figure}[t!]
\begin{center}
\includegraphics[width=0.85\textwidth, angle=0]{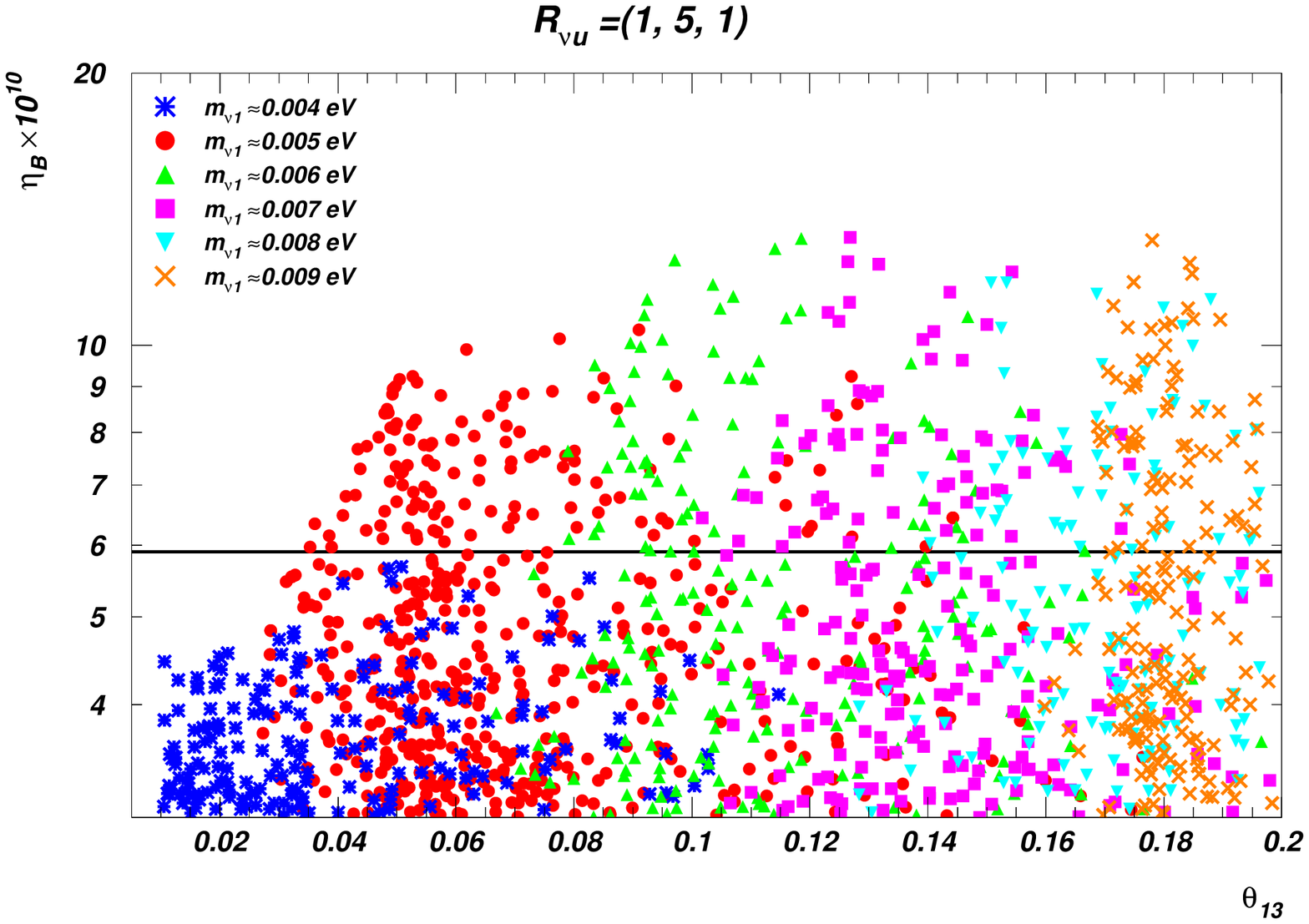}
\caption{Baryon asymmetry from thermal leptogenesis (thermal initial
$N_2$-abundance) vs. $\theta_{13}$ for different values of the
lightest neutrino mass $m_{\nu 1}$. The horizontal line marks the $2\sigma$ lower
bound, $\eta_B>5.9\times 10^{-10}$.} \label{fig:2}
\end{center}
\end{figure}

To obtain more points with sufficiently large values of
the baryon asymmetry, we now focus on a restricted
parameter space. In Fig.~\ref{fig:2}, we show results for $\eta_B$ in the
 parameter space, $2<\phi_1<4$, $-1<\phi_2<1$, $1<\delta<5$ and
$0<\theta_{13}<0.2$, with $m_{\nu 1}$ varied.
We obtain many allowed points, most of which have in common that the
asymmetry is produced in the $e$ flavor, and where
$K_{2e}+K_{2\mu}\sim 1/2$, in which case the efficiency factor is
close to maximal, and depends mildly on the initial conditions. With
a vanishing initial RH neutrino abundance the efficiency factor
would be lower by a factor of 2--3 for these points, so that the
final asymmetry would be slightly lower than that observed. Note
that although the $C\!P$ asymmetry $\varepsilon_{2\tau}$ is
typically the largest one (and so is $K_{2\tau}$), the asymmetry in
the tau flavor typically suffers from a large washout from $K_{2\tau}$;
the washout from $N_1(\tilde{N}_1)$, although set by a relatively
small $K_{1\tau}\lesssim 20$, can also have a large impact due to
its exponential effect [see Eq.~(\ref{thermal})].

From Fig.~\ref{fig:2}, we find a lower bound on
the lightest neutrino mass and on
$\theta_{13}$:
\begin{eqnarray}
m_{\nu 1}&\gtrsim &0.004~{\rm eV}\,, \ \ \ \ \ \ \ \ \ \theta_{13} \gtrsim 0.04 \,.\label{thetabound}
\end{eqnarray}
Our lower bound on the lightest neutrino mass is slightly more
restrictive than found in Ref.~\cite{DiBari:2008mp}, where a
non-SUSY framework and a vanishing RH neutrino abundance were
considered. On the other hand, our lower bound on $\theta_{13}$
agrees well with that of Ref.~\cite{DiBari:2008mp} in the mass region below
$m_1=0.009$~eV. Note that the Daya Bay~\cite{DayaBay} and Double Chooz~\cite{Chooz}
reactor experiments are sensitive to $\theta_{13}$ for $\theta_{13} \gtrsim 0.05$. 

As mentioned above, with a vanishing initial
$N_2~(\tilde{N}_2)$-abundance it is very difficult to obtain a
baryon asymmetry in the allowed range. At this point, it is worth
commenting on the size of the theoretical errors in the computation
of the baryon asymmetry. First, we have neglected spectator
processes~\cite{Buchmuller:2001sr}, including
flavor mixing in the so-called $C$ matrix~\cite{Barbieri:1999ma},
which can reduce the final asymmetry by up to
30\%~\cite{Blanchet:2008pw}. We also neglected quantum
statistical factors and assumed that kinetic equilibrium holds, an
approximation that is very good in the strong washout regime
($K_{1\alpha}\gg 1$), but which can make a 50\% difference in the
weak washout regime ($K_{1\alpha}\ll
1$)~\cite{Basboll:2006yx}. In our study, we
obtained the largest asymmetries when $K_{1\alpha}\sim 1$, in which
case the uncertainty is less than 10\%. Finally, we did not solve the
full quantum Boltzmann equations based on the Keldysh-Schwinger
non-equilibrium formalism; for recent related work see 
Ref.~\cite{Garny:2009qn}.
Note that in this formalism, thermal corrections are automatically taken into account. 
According to Ref.~\cite{Garny:2010nj} an enhancement of
the $C\!P$ asymmetry parameter by a factor of a few is possible.

It is difficult to estimate the cumulative effect of all these theoretical
uncertainties on our results.
Nevertheless, a 50\% uncertainty in the final asymmetry seems to be a
fair assessment and we conclude that a vanishing initial RH neutrino abundance
in this framework cannot be excluded.

\begin{figure}[t!]
\begin{center}
\includegraphics[width=0.85\textwidth, angle=0]{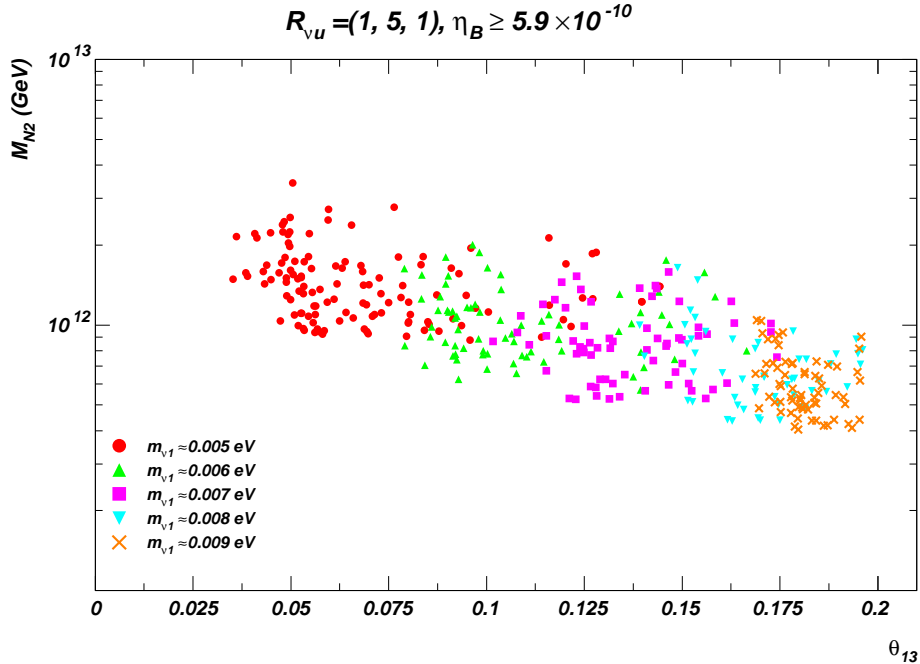}
\caption{Values of $M_{N_2}$ for which thermal leptogenesis is
successful for different values of the lightest
neutrino mass $m_{\nu 1}$.} \label{fig:BoundMN2}
\end{center}
\end{figure}

Next, we extract from Fig.~\ref{fig:2} the values of the RH neutrino
masses $M_{N_2}$ required for successful thermal leptogenesis. The
results are shown in Fig.~\ref{fig:BoundMN2}. These clearly point to
a very high reheat temperature, $T_{\rm R}\sim M_{N_2}\sim
10^{12}$~GeV, which poses a problem as we explain in the next
subsection.

\TABLE{ \centering
\begin{tabular}{l|c|c}
\hline
  & Present & Future  \\
\hline
BR($\mu \rightarrow e\gamma$)    & $1.2 \times 10^{-11}$~\cite{mega} & $10^{-13}$~\cite{meg}\\
BR($\tau \rightarrow \mu\gamma $)& $4.5 \times 10^{-8}$~\cite{belle} & $10^{-9}$~\cite{Bona:2007qt} \\
BR($\tau \rightarrow e\gamma$)   & $3.3 \times 10^{-8}$~\cite{babar} & $10^{-9}$~\cite{Bona:2007qt} \\
BR($\mu \rightarrow eee$)        & $1.0 \times 10^{-12}$~\cite{sindrum} & $10^{-14}$~\cite{Marciano:2008zz}\\
BR($\tau \rightarrow \mu\mu\mu $)& $3.2 \times 10^{-8}$~\cite{belle_3l} & $10^{-9}$~\cite{Bona:2007qt} \\
BR($\tau \rightarrow eee$)       & $3.6 \times 10^{-8}$~\cite{belle_3l} & $10^{-9}$~\cite{Bona:2007qt} \\
CR($\mu\,\textrm{Ti}\rightarrow e\,\textrm{Ti}$) & $4.3 \times
10^{-12}$~\cite{sindrum2} &
$10^{-18}$~\cite{prime}\\
CR($\mu\,\textrm{Al}\rightarrow e\,\textrm{Al}$) & - & $10^{-16}$~\cite{mu2e} \\
\hline
\end{tabular}
\caption{Present bounds and projected sensitivities for LFV
processes.}\label{tab:lfv}} We now turn to LFV. It has been
known for a long time that the supersymmetric seesaw potentially
leads to large LFV rates, due largely to slepton contributions at the
loop level~\cite{Borzumati:1986qx}. It is then a quantitative
question to know if the points compatible with thermal leptogenesis
in our framework lead to observable rates in future
experiments. We show in Table~\ref{tab:lfv} the current bounds and
projected sensitivities for LFV.

So far our results have been essentially independent of the
region of mSUGRA parameter space compatible with the dark matter
abudance since leptogenesis
occurs at very high energy scales. However, for the LFV rates, it is of
crucial importance. 
We focus on the bulk region, which is the most optimistic region 
for the detection of LFV.
This region is characterized by the following
parameters: $m_0=80$~GeV, $m_{1/2}=170$~GeV, $A_0=-250$~GeV and
$\tan\beta=10$. We also checked that points in the stop-coannihilation region 
($m_0=150$~GeV, $m_{1/2}=300$~GeV, $A_0=-1095$~GeV and $\tan\beta=5$) 
yield rates that are no more than a factor of
two different than in the bulk region~\cite{Barger:2009gc}. This is not
surprising given the fact that LFV rates are maximized when
sfermions and gauginos are light, and when the mass hierarchy
is mild between them.

\begin{figure}[t!]
\begin{center}
\includegraphics[width=0.85\textwidth, angle=0]{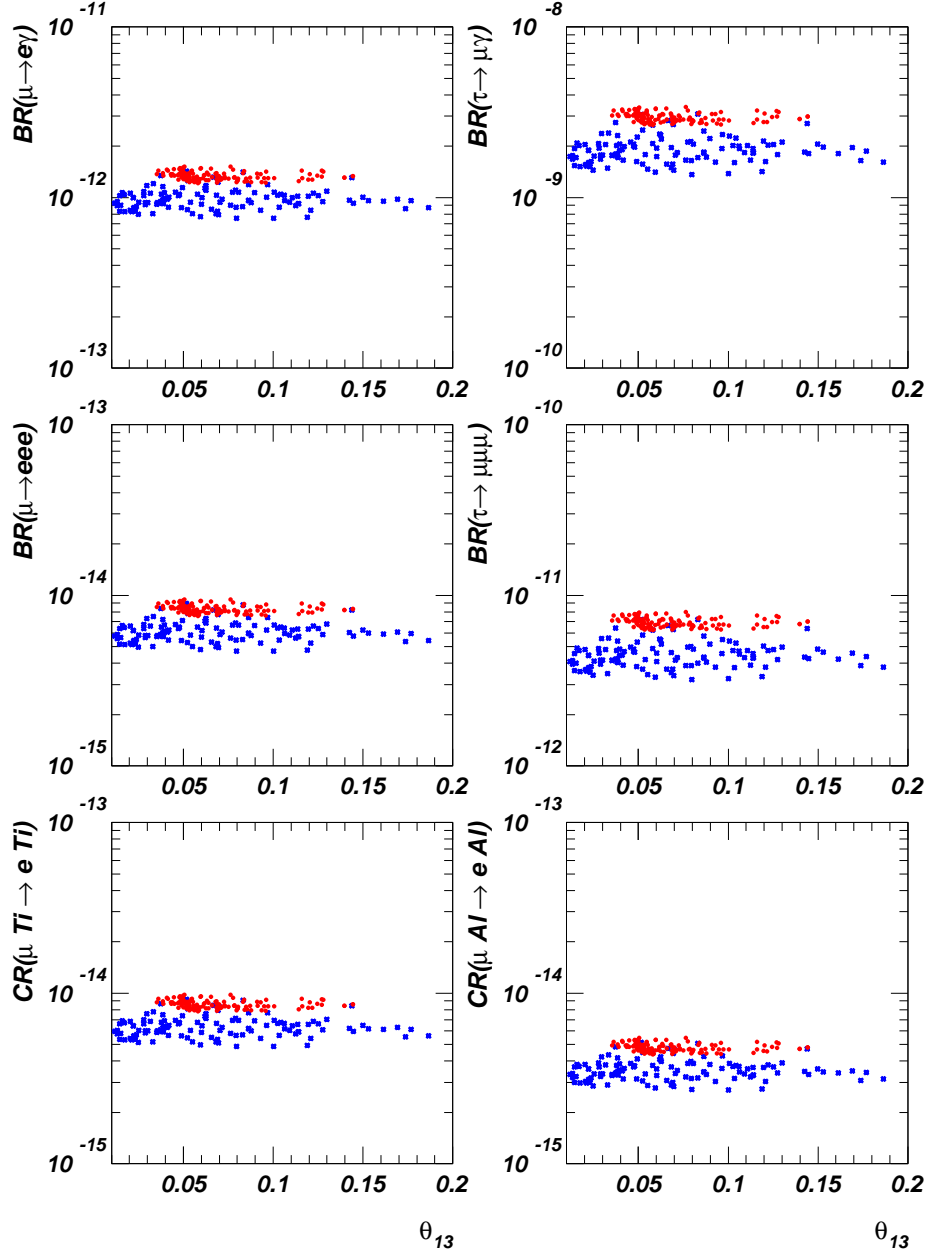}
\caption{LFV rates for points with an asymmetry above $5.9\times
10^{-10}$ within thermal leptogenesis (red dots) and
non-thermal leptogenesis at preheating (blue stars).}
\label{fig:3}
\end{center}
\end{figure}

The results for LFV rates in the bulk region are presented in
Fig.~\ref{fig:3}.
We find that the allowed points lead to predictions
for LFV rates below the present exclusion limits, and, except
for $\tau\to \mu\mu\mu$, within reach of current and future
experiments. In particular, MEG should see a $\mu \to e\gamma$ signal if leptogenesis
is realized in our framework. It should be noted that LFV rates
are very weakly dependent on the value of the lightest neutrino mass
$m_{\nu 1}$, and we checked that the rates are essentially
unchanged in the range \mbox{$0.004~{\rm eV}<m_{\nu 1}<0.009$~eV}.

We end this section by commenting on the dependence of LFV rates on $R_3$. 
Increasing $R_3$ leads to a larger $M_{N_3}$ [see
Eq.~(\ref{eq:ckmRHN})], which in turn leads to larger Yukawa
couplings $(\bfn)_{\alpha 3}$ in order to keep the neutrino mass
matrix fixed. We expect LFV rates to increase as we increase
$R_3$ since the third generation Yukawa contributes dominantly to the rates. We explicitly checked that this is the case, and found
the rates for $R_3=5$ to be about one order of magnitude larger than those shown in
Fig.~\ref{fig:3}.

\subsection{Reheat temperature and the gravitino problem}

In order for leptogenesis to work in our framework, we need a very
high reheat temperature ($\sim 10^{12}$ GeV) of the Universe after inflation 
(see Fig.~\ref{fig:BoundMN2}). Therefore, the gravitino
problem here is even more severe than in the conventional scenario,
where typically a reheat temperature of about $10^9$~GeV is
sufficient.

Let us first consider the case where the gravitino is \emph{not} the
lightest supersymmetric particle (LSP). The gravitino problem arises
 because the thermal production of gravitinos is unavoidable
after inflation. The gravitino yield, $Y_{3/2}$, defined as the
gravitino number density divided by the entropy density, is nearly
proportional to the reheat temperature~\cite{Kawasaki:2008qe}:
\begin{eqnarray}
Y_{3/2}&\simeq& 2.3\times 10^{-14}\times T_{\rm
R}^{(8)}\left(1+0.015~\log T_{\rm R}^{(8)}-0.0009 \log^2T_{\rm
R}^{(8)}\right)\nonumber\\
&&+1.5\times 10^{-14}\times \left({m_{1/2}\over m_{3/2}}\right)^2
T_{\rm R}^{(8)}\left(1-0.037 \log T_{\rm R}^{(8)}+0.0009
\log^2T_{\rm R}^{(8)}\right)\, ,\label{gravitino}
\end{eqnarray}
where $T_{\rm R}^{(8)}\equiv T_{\rm R}/(10^8$~GeV), $m_{3/2}$ is the
gravitino mass, and $m_{1/2}$ is the unified gaugino mass at the GUT
scale. Note also that increasing the
gravitino mass lowers its abundance until the first term in
Eq.~(\ref{gravitino}) dominates, at
which point the abundance saturates.

Gravitinos being only gravitationally coupled, for masses below
30~TeV they typically decay during or after Big Bang Nucleosynthesis (BBN), 
hence spoiling the agreement between observations
and theory. Requiring that BBN is successful leads
to stringent constraints on the reheat temperature as a function of
the gravitino mass (see~\cite{Kawasaki:2008qe} and references
therein): $T_{\rm R}<10^6$~GeV if the gravitino mass is lower than about 10~TeV.

For $m_{3/2}\gtrsim 30$~TeV, the bound relaxes to
$T_{\rm R}\lesssim 10^{10}$~GeV so that
that the gravitino decays into the LSP (assumed to be
the lightest neutralino) yielding a non-thermal contribution
that saturates the observed dark matter density. This upper bound is
still at odds with thermal leptogenesis in our framework.
Therefore, we conclude that thermal leptogenesis is somewhat incompatible
with neutralino dark matter.

Suppose we abandon neutralino dark matter. Consistent
cosmology first requires that the gravitino be heavier than 30~TeV
so that it decays to the LSP before the onset of BBN. In turn, the LSP may decay
to a hidden sector~\cite{DeSimone:2010tr}. The lightest hidden
sector particle $X$ needs to be much lighter than the LSP so that
its energy density is diluted before matter-radiation equality:
\begin{equation}
\Omega_X h^2 \simeq 2.8\times 10^{5}\times Y_{3/2} \left({m_X\over
1~{\rm MeV}}\right).
\end{equation}
Using Eq.~(\ref{gravitino}) for a very heavy gravitino $m_{3/2}\gg
m_{1/2}$, we find
\begin{equation}
\Omega_X h^2 \simeq 6\times 10^{-5} \,\left({T_{\rm R}\over
10^{12}~{\rm GeV}}\right)\,\left({m_X\over 1~{\rm MeV}}\right)\,.
\end{equation}
The hidden sector particle could constitute part of the
dark matter, but being warm, its abundance cannot exceed 5\% of
the total dark matter abundance~\cite{Boyarsky:2008xj}.

An alternative way to circumvent the gravitino problem is if the gravitino is the
(visible) LSP, with a mass of about 200--300~GeV, and itself decays
into a much lighter hidden sector particle long before
matter-radiation equality. In this case, the NLSP must decay before
BBN, which can be easily achieved with  hidden
sector dynamics~\cite{DeSimone:2010tr}. Note that the gravitino cannot be
the dark matter particle.

Independently of $m_{3/2}$, we find that dark matter must be explained by
an external mechansim and particle, like the axion.


\section{Non-thermal leptogenesis at preheating}

As explained in the previous section, within our mSUGRA-seesaw
framework with $SO(10)$-inspired boundary conditions, standard
thermal leptogenesis is subject to a severe gravitino problem.

Non-thermal leptogenesis allows for a low reheat temperature so that
neutralino dark matter is viable. Non-thermal RH
neutrino production can be obtained either during the preheating
stage~\cite{Giudice:1999fb}, or from inflaton
decays~\cite{Asaka:1999jb,Asaka:1999yd}.

Here we follow the approach of Ref.~\cite{Giudice:2008gu} which does not
require any additional ingredient to our framework such as a large coupling of the RH neutrinos to the
inflaton. The idea is to use the presence of flat directions in the
scalar potential (for a review see~\cite{Enqvist:2003gh}) to enable
instant preheating~\cite{Felder:1998vq}, which is a very efficient
way of producing very heavy states. We briefly review the mechanism.

$F$- and $D$-term flat
directions can be lifted by soft supersymmetry (SUSY) breaking terms in our
vacuum, with non-renormalizable terms in the superpotential, or with
finite density terms in the potential~\cite{Dine:1995uk}. If the
soft SUSY-breaking term has a positive sign and the finite density
term (proportional to the Hubble rate squared $H^2$), contributes
negatively, the field $\phi$ along the flat direction initially acquires
a large vacuum expectation value (VEV) denoted by $\phi_0$.
After inflation ends, the inflaton starts oscillating at the
minimum of its potential, while the Hubble rate falls. Once
$H\sim \widetilde{m}/3$, where $\widetilde{m}$ is a soft
SUSY-breaking mass term, the flat direction starts moving down
towards the true minimum at $\phi=0$.
Following~\cite{Giudice:2008gu}, we assume that the condensate
involves the third generation quark $u_3$, and focus on the
production of the up-type scalar Higgs $H_u$ relevant for
leptogenesis. The condensate couples to the Higgs through the term
$f_t|\phi|^2|H_u|^2$. If the condensate passes through the origin
(or sufficiently close to it), Higgses will be produced when
adiabaticity is violated~\cite{Felder:1998vq}, i.e.
$\dot{m}_{H_u}/m_{H_u}^2\gtrsim 1$.

The condensate continues its motion upwards after it has passed
through the origin, and the up-type Higgs effective mass 
gradually increases proportionally to $f_t|\phi|$. When the Higgs
effective mass becomes larger than the RH neutrino mass, it 
promptly decays to RH neutrinos. The heaviest particles that can be
produced through this mechanism have a mass~\cite{Giudice:2008gu}
\begin{equation}
M^{\rm max}\simeq 4\times 10^{12}~{\rm GeV} \,\left({|\phi_0|\over
M_{\rm Pl}}\right)^{1/2}\, \left({\widetilde{m}\over 100~{\rm
GeV}}\right)^{1/2} \,,
\end{equation}
which is large enough for our purposes. Accounting for the fact that
reheating occurs after leptogenesis, a large dilution factor must be
included. Assuming that all up-type Higgses decay into RH neutrinos, the
final result for the baryon asymmetry is given
by~\cite{Giudice:2008gu}
\begin{equation}\label{eqpreh}
\eta_{B,i}\simeq 7\times 10^{-5}\, \sum_{\alpha}\tilde{\varepsilon}_{i\alpha} \, \left({T_{\rm
R}\over 10^8~{\rm GeV}}\right)\,\left({|\phi_0|\over M_{\rm
Pl}}\right)^{3/2}\,\left({100~{\rm GeV}\over
\widetilde{m}}\right)^{1/2}\,,
\end{equation}
which means that for the canonical choice of parameters the efficiency factor
is about $7\times 10^{-3}$.

For a highly hierarchical RH neutrino mass spectrum, and small
$M_{N_1}$, we need to again consider the production of
asymmetry from $N_2$. Specifically, we need the up-type Higgs to
promptly decay into $N_2$ rather than $N_1$, which is possible
if~\cite{Giudice:2008gu}
\begin{equation}
M_{N_2}\lesssim \left({8\pi f_t \widetilde{m} |\phi_0|\over
\sum_{\alpha} |({\bf f}_\nu)_{\alpha 1}|^2}\right)^{1/2}\,.
\end{equation}
The condition can be easily satisfied if $|({\bf f}_\nu)_{\alpha 1}|\ll 1$.

Finally, we need to take into account the washout by $N_1$ inverse
decays of the asymmetry produced by $N_2$. Using Eq.~(\ref{eqpreh})
we find
\begin{equation}\label{asympreheating}
\eta_{B,2}\simeq 7\times 10^{-5}\, \sum_{\alpha}
\tilde{\varepsilon}_{2\alpha}\exp\left(-{3\p\over
4}K_{1\alpha}\right)\,,
\end{equation}
where we have taken $T_{\rm R}=10^8$~GeV, $\widetilde{m}=100$~GeV and
$\phi_0=M_{\rm Pl}$.

Before turning to the numerical results, we comment on
another non-thermal scenario of leptogenesis, namely at reheating
from inflaton
decay~\cite{Asaka:1999jb,Asaka:1999yd}. The final asymmetry 
obtained for an inflaton of mass $10^{13}$~GeV is~\cite{Asaka:1999jb}
\begin{equation}
\eta_{B,i}\simeq {7\over 2}\times 10^{-5} \,\sum_{\alpha}\tilde{\varepsilon}_{i\alpha}\,{\rm
Br}(\phi\to N_iN_i)\,\left({T_{\rm R}\over 10^8~{\rm GeV}}\right)\, ,
\end{equation}
where ${\rm Br}(\phi\to N_iN_i)$ is the branching ratio of the inflaton
decay channel $\phi \to N_iN_i$. We find that even if the branching ratio
to the next-to-lightest RH neutrino $N_2$ is unity, the maximum
efficiency factor is a factor of 2 smaller than in the preheating case.
All the results presented in the next subsection can then be
trivially extended to the reheating case.

\subsection{Results}

\begin{figure}
\center
\includegraphics[width=0.85\textwidth]{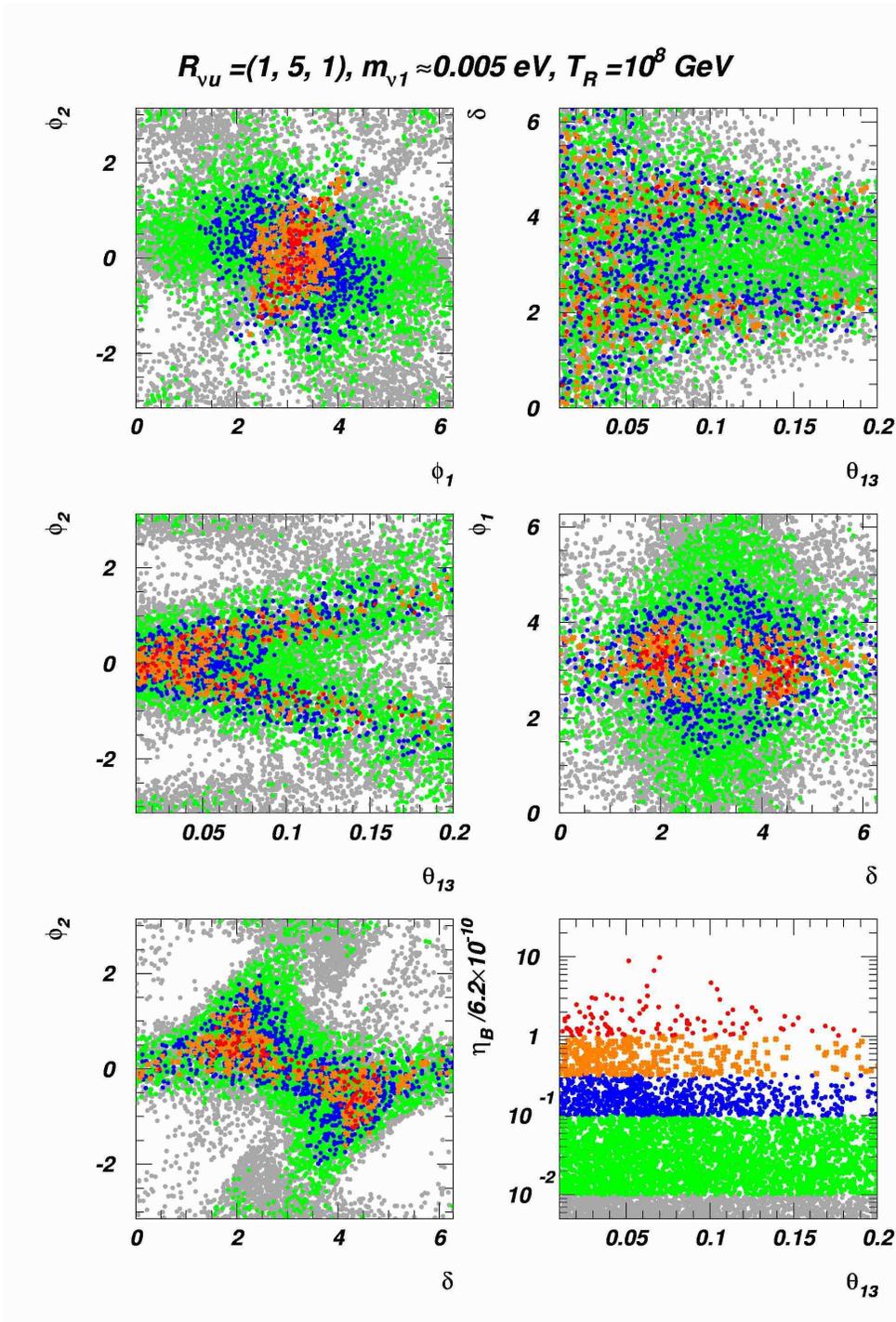}
\caption{Full parameter space scan for non-thermal leptogenesis at
preheating. The color code is evident from the bottom-right
panel.} \label{fig:4}
\end{figure}

Similarly to the previous section, we explore the
parameter space for which Eq.~(\ref{asympreheating}) reproduces the
observed baryon asymmetry; see Fig.~\ref{fig:4}.
Note that the measured baryon asymmetry can be generated
for more of the parameter space than with thermal
leptogenesis. The reason is simply that the asymmetry depends 
on fewer paremeters than in the thermal case. In particular, the asymmetry
does not depend on $K_{2\alpha}$ as was the case for
thermal leptogenesis [see Eqs.~(\ref{asympreheating})
and~(\ref{thermal})]. Therefore, non-thermal leptogenesis 
is possible even for very large values of the washout parameter $K_{2\alpha}$, 
so long as $K_{1\alpha}$ remains small. In the non-thermal case, the $SO(10)$ mass
relations in Eq.~(\ref{quarkneutrino}) and light neutrino masses  
lead us to a new part of the
parameter space in which the Yukawa coupling $({\bf f}_\nu)_{\tau 2}$ 
(and therefore $K_{2\tau}$) is large, while $K_{1\tau}$
remains small. In this region, the lepton asymmetry is produced in
the $\tau$ flavor because of the large value of the $C\!P$ asymmetry
parameter $\varepsilon_{2\tau}$ and can be up to two orders of
magnitude larger than the observed value.
Due to the greater freedom in the choice of parameters, it is not
surprising that neither
$m_{\nu 1}$ nor $\theta_{13}$ are bounded from below.

For the points in Fig.~\ref{fig:4}, we show the
corresponding LFV rates in Fig.~\ref{fig:3}. As for thermal
leptogenesis, we find that large rates are predicted for the bulk and
stop-coannihilation regions. In
particular, MEG should see a positive signal if leptogenesis is 
the origin of the baryon asymmetry. Note that the
rates remain essentially unchanged under variations of the lightest neutrino
mass $m_{\nu 1}$ within the range of interest.

Finally, let us comment on the gravitino problem in this framework.
Since the reheat temperature after inflation is required to be of order
$10^7$--$10^8$~GeV, it is clearly not as severe as in the previous
section. However, if dark matter is to be explained by the standard
neutralino LSP, we still need a fairly heavy gravitino of about
10~TeV~\cite{Kawasaki:2008qe}. On the other hand, if the gravitino
is the LSP, we must ensure that the NLSP decays before BBN, which
can be easily achieved with small R-parity
violation~\cite{Buchmuller:2007ui}, or with decays into a light
hidden sector~\cite{DeSimone:2010tr}.

\section{Conclusions}
\label{sec:conclus}

We studied the implications of successful leptogenesis on the
mSUGRA-seesaw parameter space with dark matter comprised of neutralinos. 
Guided by $SO(10)$-inspired mass relations, we were led to
hierarchical Dirac mass eigenvalues in the
neutrino sector.

We found that with thermal leptogenesis, a large enough baryon
asymmetry is difficult to realize, and obtained lower bounds on
both the lepton mixing angle $\theta_{13}$ and the lightest neutrino
mass $m_{\nu 1}$. The LFV rates in the bulk and stop-coannihilation
regions are large and observable at
current experiments such as MEG. However, the high reheat
temperature \mbox{($\sim 10^{12}$~GeV)}
implies a severe overproduction of gravitinos, rendering
neutralino dark matter and thermal leptogenesis 
somewhat incompatible in our framework.

To relax the tension with gravitino overproduction, so as to not
abandon neutralino dark matter, we explored the possibility of non-thermal
leptogenesis in the preheating phase which relies on the
mechanism of instant preheating from supersymmetric flat
directions already present in our model. We found that an
efficient (non-thermal) production of RH neutrinos can be achieved,
even for reheat temperatures as low as $10^7$~GeV. Note that
non-thermal leptogenesis from inflaton decay would lead to very
similar results. The parameter space for successful leptogenesis at
preheating is less constrained than in the thermal framework, and
LFV rates in the bulk and stop-coannihilation regions
 remain large and observable at current and future
experiments.

\section*{Acknowledgments}
We thank Z.~Chacko, A.~Ibarra and A.~Kartavtsev for
useful discussions. This work was
supported by the DoE under Grant Nos. DE-FG02-04ER41308 and DE-FG02-94ER40823, 
by the NSF under Grant No. PHY-0544278, and by the Maryland Center for Fundamental Physics.


\end{document}